\documentclass{cjaa}                    % preprint, the final version for publication

\usepackage{graphicx}                   %for PS/EPS graphics inclusion, new
\input{epsf.sty}                        %for PS/EPS graphics inclusion, old
\input{psfig.sty}                       %for PS/EPS graphics inclusion, old

\begin{document}

   \title{TeV Gamma-Ray Astrophysics
%\footnotetext{$*$ Supported by the National Natural Science Foundation of China.}
}
%   \subtitle{I. Place Your Subtitle Here}

%   \baselineskip=5mm     %%preserved for Editor. DOn't remove!

   \author{Marc Rib\'o
      \inst{}\mailto{mribo@am.ub.es}
%% Please move "\mailto{}" to the corresponding author of the paper
%% For single author or all the authors from an institute, use "\inst{}"
%   \and E. Rodr\'{\i}guez
%      \inst{2}
%   \and M. Breger
%      \inst{3}
      }
   \offprints{Marc Rib\'o}                   %% is disabled in fact

   \institute{Departament d'Astronomia i Meteorologia, Universitat de Barcelona, Mart\'{\i} i Franqu\`es 1, E-08028 Barcelona, Spain\\
             \email{mribo@am.ub.es}
%% offprint requests will be sent. Note to pair \mailto{} with \email{}
%        \and
%             Instituto de Astrofisica de Andalucia, CSIC,
%             P. O. Box 3004, E-18080 Granada, Spain\\
%             \email{eloy@iaa.es}
%        \and
%             Astronomisches Institut der Universit\"{a}t Wien, T\"{u}rkenschanzstr. 17,
%             A-1180 Wien, Austria\\
%             \email{breger@astro.univie.ac.at}
          }

   \date{Received~~2007 September 28; accepted~~2007 November 25}

   \abstract{
The window of TeV Gamma-Ray Astrophysics was opened less than two decades ago,
when the Crab Nebula was detected for the first time. After several years of
development, the technique used by imaging atmospheric Cherenkov telescopes
like HESS, MAGIC or VERITAS, is now allowing to conduct sensitive observations
in the TeV regime. Water Cherenkov instruments like Milagro are also providing
the first results after years of integration time. Different types of
extragalactic and galactic sources have been detected, showing a variety of
interesting phenomena that are boosting theory in very high energy gamma-ray
astrophysics. Here I review some of the most interesting results obtained up to
now, making special emphasis in the field of X-ray/gamma-ray binaries.
%   \keywords{techniques: photometric -- stars: variables: $\delta$ Scuti --
%   stars: individual: AD Ari, IP Vir, YZ Boo  }
   }

%   \authorrunning{M. Rib\'o }            %author_head in even pages
%   \titlerunning{TeV Gamma-ray Astrophysics }  % title_head in odd pages

%% The author head (on even pages) and the title head (on odd pages) will be
%% automatically extracted from \author and \title. Whenever the title is too long,
%% you will be asked to supply a shorter one by inserting either \authorrunning{} or
%% \titlerunning{} before \maketitle. Anyway, you can specify your own heads.

   \maketitle
%
%________________________________________________ sections below
%
%% ChJAA editors DO NOT use \cite{} for citation, \ref and \label for
%% cross-references of Table/Figure in publication version, so ChJAA prefers you giving
%% the year as Michel et al. 1992, and writting Table~1 or Fig.~1 and so forth.
%% However, that will make authors inconvenient in adjusting/adding/removing
%% text/table/figure.
%%
\section{Introduction}
%% first-level sections will be auto-capitalized
\label{sect:intro}
%\hspace{15pt}%                   %% preserved for Editor

There are different astrophysical scenarios in which particles are accelerated
up to TeV energies. Interestingly, these particles can produce TeV gamma-ray
photons that can travel in straight lines from the original sources to the
observer. Therefore, observational TeV astronomy can provide very useful
information to constrain astrophysical scenarios for particle accelerators.

Although the idea to detect gamma-rays of TeV energies using the Cherenkov
light that they produce when entering the atmosphere was introduced about 50
years ago, it took a few decades of effort to clearly detect the first source,
the Crab Nebula in 1989. After a few years of observations with the third
generation of imaging atmospheric Cherenkov telescopes, about 75 TeV sources
have been detected. Gamma-ray observations in the TeV regime can certainly be
considered nowadays as a new branch of observational astronomy.

Here I review the beginning of TeV Gamma-Ray Astrophysics
(Sect.~\ref{sect:history}), the current instrumentation
(Sect.~\ref{sect:instrumentation}), the most interesting results obtained up to
now (Sect.~\ref{sect:results}), and make special emphasis in the field of
X-ray/gamma-ray binaries (Sect.~\ref{sect:binaries}). I finish with some
comments on future instrumentation (Sect.~\ref{sect:future}) and with the
conclusions (Sect.~\ref{sect:conclusions}).

\section{Historical Background}
\label{sect:history}
%\hspace{15pt}%                   %% preserved for Editor

A detailed description of the early history of the atmospheric Cherenkov 
technique can be found in Weekes (\cite{weekes05,weekes06}), while interested
readers can find an extensive review on gamma-ray astronomy with imaging
atmospheric Cherenkov telescopes in Aharonian \& Akerlof (\cite{aharonian97}).
Here I summarize the most significant facts to provide the reader an historical
background.

The idea that perhaps about 0.01\% of the night-sky photons were the result of
Cherenkov light emitted by cosmic rays and secondary particles entering the
atmosphere of the Earth was introduced by Blackett (\cite{blackett49}). A few
years later, Galbraith \& Jelley (\cite{galbraith53}) built a rudimentary
instrument with a 25\,cm mirror inside a garbage can and a photomultiplier, and
detected light-pulses of short duration correlated with cosmic radiation
detected by an air shower array.

Cherenkov light from an air shower is emitted within a cone of $\sim$120\,m
radius at ground level. Therefore, a receiver of modest dimensions has a huge
effective area. Moreover, the light pulses preserve the direction of the
primary particle, and the amount of light is proportional to the energy of the
primary particle. Therefore, this technique could be used to do gamma-ray
astronomy at Very High Energies (VHE).

During the 60s two experiments using searchlight mirrors from the second world
war were built in Crimea (1960--1965) and Ireland (1962--1966). The Whipple
10\,m telescope, inaugurated in 1968 and still working, was the first large
optical reflector built for gamma-ray astronomy. All these instruments were
first generation Cherenkov telescopes in which no gamma-ray and hadronic
primary differences were considered in the analysis. The background of air
showers produced by cosmic rays is so high compared to TeV gamma rays (a factor
between 1000--10000) that, even if cosmic rays are isotropic, no astrophysical
sources of VHE gamma rays were detected. It was clear that the technique had to
be improved to get rid of as much as cosmic ray showers as possible.

A possible solution to the gamma/hadron separation problem was first pointed
out by Jelley \& Porter (\cite{jelley63}), who enumerated several advantages of
using Cherenkov light images (including stereoscopy). The images could be
produced if the single photomultiplier placed in the focus was replaced by an
array of photomultipliers in the focal plane of the telescope (Weekes \& Turver
\cite{weekes77}). There are several factors that cause the observed shape and
size of a Cherenkov image: not only the nature of the primary, its energy and
trajectory, but also the physical process in the particle cascade, Coulomb
scattering of shower electrons, geomagnetic deflections, the distance of the
shower core from the optic axis, the Cherenov angle of emission, and the effect
of atmospheric absorption, together with the factors related to the focusing
system and the detector itself. All of them have to be considered when trying
to estimate the nature and original energy and trajectory of the particle.

After performing detailed Monte Carlo simulations of the air showers, Hillas
(\cite{hillas85}) found that the images produced by gamma rays (similar to
ellipses) would be in most cases significantly different than those produced by
cosmic rays. Moreover the fitted ellipses would point towards the center of the
image and provide further information that could be used to characterize the
showers. These developments finally allowed to detect the Crab Nebula at
9.0$\sigma$ with the Whipple 10\,m telescope equipped with a 37-pixel camera on
the focal plane (Weekes et al. \cite{weekes89}). This was the first unambiguous
detection of a TeV gamma-ray source, provided by a second generation Cherenkov
telescope.

The HEGRA experiment, with 5 telescopes of 3.4\,m diameter each, demonstrated
the advantages of stereoscopic observations using an array of telescopes
(P\"uhlhofer et al. \cite{puhlhofer03}). By observing Cherenkov images of
gamma-ray showers simultaneously with more than one telescope, one can
determine the direction of gamma-rays accurately as crossing points of image
axes. Therefore, one can explore the spatial structure of the gamma-ray objects
as well as discriminating background cosmic-ray showers more easily. Moreover,
the energy resolution of gamma-rays is improved knowing the production height
of Cherenkov light. All these facts improve the background rejection and the
sensitivity of the Imaging Atmospheric Cherenkov Telescopes (IACTs). Indeed,
HEGRA achieved an angular resolution of 0.1$^{\circ}$ for observations above
0.5\,TeV, with an energy resolution of about 15\% and a sensitivity of
$\sim$3\% of the Crab Nebula flux in 100\,h of observations. The IACTs of third
generation use (or will use) arrays of telescopes, of more than 10\,m diameter
each, working stereoscopically.

\section{Current Instrumentation}
\label{sect:instrumentation}
%\hspace{15pt}%    %% preserved for Editor

There are currently four IACTs of third generation:

\begin{itemize}

\item CANGAROO-III (Collaboration of Australia and Nippon for a GAmma Ray
Observatory in the Outback)\footnote{\tt
http://icrhp9.icrr.u-tokyo.ac.jp/index.html}

\item HESS (High Energy Stereoscopic System)\footnote{\tt
http://www.mpi-hd.mpg.de/hfm/HESS/}

\item MAGIC (Major Atmospheric Gamma Imaging Cherenkov)\footnote{\tt
http://wwwmagic.mppmu.mpg.de/}

\item VERITAS (Very Energetic Radiation Imaging Telescope Array
System)\footnote{\tt http://veritas.sao.arizona.edu/}

\end{itemize}

Their main properties are listed in Table~\ref{table:iacts}. As it can be seen,
CANGAROO-III and HESS are located in the Southern Hemisphere, while MAGIC and
VERITAS are placed in the Northern Hemisphere. This allows for a double check
of some of the most southern or northern sources, always welcome in TeV
astrophysics. CANGAROO-III, due to its low elevation, has a high threshold of
about 500\,GeV. The other instruments can reach the 100--200\,GeV regime and
even lower in the case of MAGIC, thanks to its huge dish and the high elevation
of the site. All these IACTs have to be pointed to the specific regions of the
sky to be studied.

%-----------------------------------------------------------------------------
\begin{table}[]
\caption[]{Third Generation of Imaging Atmospheric Cherenkov Telescopes (IACTs).}
%%Please Capitalize the First Letter of Each Notional Word in table's caption
\label{table:iacts}
\begin{center}
\begin{tabular}{llllllll}
\hline\noalign{\smallskip}
IACT & Location & Elevation & Telescopes & Aperture & FOV & Pixels/ & Energy \\
     &          & [km]      &        & [m] & [$^{\circ}$] & camera  & [GeV] \\
\hline\noalign{\smallskip}
CANGAROO-III & Woomera, Australia & 0.2 & 4      & 10      & 4.0 & 427 & $>$500 \\
HESS         & Gamsberg, Namibia  & 1.8 & 4 (+1) & 12 (28) & 5.0 & 960 & $>$100 \\
MAGIC        & La Palma, Spain    & 2.2 & 1 (+1) & 17 (17) & 3.5 & 576 & $>$~\,60 \\
VERITAS      & Arizona, USA       & 1.3 & 4      & 12      & 3.5 & 499 & $>$100 \\
\noalign{\smallskip}\hline
\end{tabular}
\end{center}
\end{table}
%-----------------------------------------------------------------------------

HESS started stereoscopic observations in 2003, MAGIC normal operations in 2004
and CANGAROO-III in 2005, while VERITAS started stereoscopic observations at
the end of 2006 and has just completed the array in 2007. The most interesting
results published up to now have been provided by HESS and MAGIC, and I will
concentrate on them. As a reference, the flux sensitivity at the 5$\sigma$
level in 50\,h of observations is 2\% of the Crab Nebula at 250\,GeV with MAGIC
and $\simeq$1\% with HESS.

In parallel to IACTs, other TeV detectors using different techniques have been
built during the last decade. Among them, the Milagro Gamma Ray
Observatory\footnote{\tt http://www.lanl.gov/milagro/;
http://umdgrb.umd.edu/cosmic/milagro.html}, a water Cherenkov extensive air
shower array, started operations in 2000, and has provided its first impressive
results during the last year. Milagro, located in New Mexico (USA), explores
the 2--150\,TeV energy range and has a FoV of 2\,sr. It can thus provide an
unbiased sky survey of the northern skies at multi-TeV energies.

\section{General results}
\label{sect:results}
%\hspace{15pt}%                   %% preserved for Editor

Note that reviews on some of the results obtained by HESS and MAGIC were also
presented in this conference and can be found in Santangelo
(\cite{santangelo07}) and Bartko (\cite{bartko07}), respectively.

\begin{figure}
\centering
\resizebox{1.0\textwidth}{!}
{\includegraphics[]{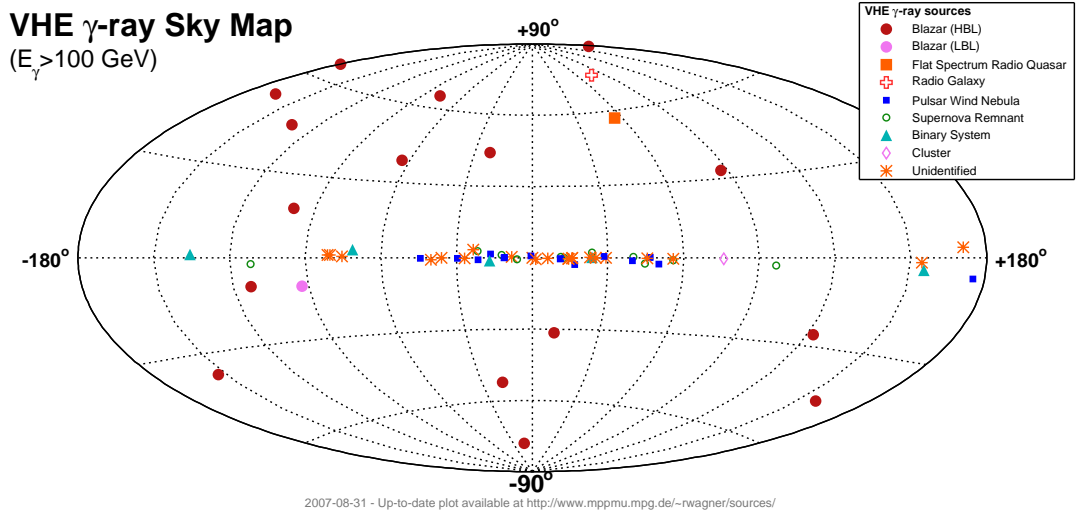}}
\caption{VHE gamma-ray sky map as of 2007 August 31 from Robert Wagners' {\tt http://www.mppmu.mpg.de/$\sim$rwagner/sources/}. 19 extragalactic sources with redshifts between 0.0044 (M87) and 0.536 (3C~279) are plotted together with 56 Galactic sources including pulsar wind nebulae, supernova remnants, X-ray binaries and many unidentified sources. Look at {\tt http://tevcat.uchicago.edu/} for an interactive web page on all TeV sources.}
\label{fig:sky}
\end{figure}

In the reviews on TeV astronomy written during the last years or decades it has
been usual to provide a list of the known sources. Since this number is about
75 sources at the time of writing (and growing!), this is not practical any
more. Instead, it is more instructive to show only the distribution of TeV
sources in the sky, in galactic coordinates, and using different symbols for
each type of object. A plot of this kind is shown in Fig.~\ref{fig:sky}
(courtesy of Robert Wagners' {\tt
http://www.mppmu.mpg.de/$\sim$rwagner/sources/}). According to these data,
there are 19 extragalactic sources and 56 galactic sources. Let us now review
the main properties of these sources and some of the physics that can be
learned from them.

%-----------------------------------------------------------------------------
%-----------------------------------------------------------------------------

\subsection{Extragalactic Sources}
\label{sect:extragalactic}
%\hspace{15pt}%                   %% preserved for Editor

{\bf High frequency peaked BL Lacertae objects (HBL)}. BL Lacertae objects are
Active Galactic Nuclei (AGN) with relativistic jets pointing close to the line
of sight. The relativistic electrons in the jets produce synchrotron radiation
from radio to X-rays thanks to the presence of magnetic fields. On the other
hand, VHE gamma-ray emission can be produced if these electrons suffer inverse
Compton (IC) scattering with ambient photons (although hadronic processes can
also be at work). These photons can be those emitted by the synchrotron
mechanism (Synchrotron Self Compton or SSC), or photons from external photon
fields (External Compton or EC). The Spectral Energy Distributions (SEDs) of BL
Lacertae objects show a characteristic double bump formed by the synchrotron
peak and the Compton peak. The objects where the synchrotron peak is placed at
high frequencies (UV to X-rays) are more easily detectable at TeV energies,
since the energy gain in the Compton process does not need to be so dramatic.
These objects are called High frequency peaked BL Lacertae objects (or HBL).
After the discovery of the first TeV source, the three following ones were the
HBLs Markarian~421 ($z=0.031$; Punch et al. \cite{punch92}), Markarian~501
($z=0.034$; Quinn et al. \cite{quinn96}) and 1ES~2344+514 ($z=0.044$; Catanese
et al. \cite{catanese98}). Since then, 13 HBL have been discovered with
redshifts between 0.047 and 0.212.

{\bf Extragalactic Background Light (EBL)}. The newly discovered high redshift
HBLs have allowed to study the attenuation of the TeV emission by photon-photon
absorption and pair creation due to Extragalactic Background Light (EBL). In
particular, the HESS detections of 1ES~1101$-$232 ($z=0.186$) and H~2356$-$309
($z=0.165$) showing very hard spectra, imply a low level of EBL (Aharonian et
al. \cite{aharonian06_ebl}). In other words, most of the light emitted at
optical/near-infrared wavelengths appears to be very close to the lower limit
given by the integrated light of resolved galaxies. This effectively excludes
contribution from other sources like the first stars formed, and shows that the
intergalactic medium is more transparent to gamma-rays than previously thought.
It is interesting to note that, based on the current observational techniques,
any possible TeV extragalactic background would not be detected (even if it is
not very absorbed).

{\bf Rapid flares in HBLs}. Rapid flares in Markarian~501 and PKS~2155$-$304
have recently been discovered by MAGIC and HESS, respectively. In the first
case, the source showed flux-doubling times down to 2~minutes and an indication
of a 4$\pm$1 min time delay between the peaks of F($<$0.25\,TeV) and
F($>$1.2\,TeV), which may indicate a progressive acceleration of electrons in
the emitting plasma blob (Albert et al. \cite{albert07_flare_mkn501}), but also
be the result of a vacuum refractive index in the context of quantum gravity
(Albert et al. \cite{albert07_qg}). In the case of PKS~2155$-$304,
flux-doubling times down to 3~minutes were measured, and assuming an emitting
region with the size of the Schwarzschild radius of a 10$^9$\,M$_\odot$ black
hole, Doppler factors above 100 are required, challenging our current
understanding of blazar jets (Aharonian et al.
\cite{aharonian07_flare_pks2155}).

{\bf Low frequency peaked BL Lacertae objects (LBL)}. LBLs have their
synchrotron peak in the submillimiter to optical bands. The first firm
detection of an LBL has recently been reported by MAGIC for the prototype
source BL Lacertae ($z=0.069$, Albert et al. \cite{albert07_bllac}) ruling out,
thanks to simultaneous optical data, previous claims of much higher detections
by the Crimean group. The multi-wavelength data favors a leptonic scenario.

{\bf The radio galaxy M87}. The Fanaroff-Riley type I radio galaxy M87
($z=0.0043$) has been detected at VHE gamma-rays by HESS (Aharonian et al.
\cite{aharonian06_m87}), confirming a previous detection reported by HEGRA. The
observed fast variations indicate that the emission comes from a region with a
dimension similar to the Schwarzschild radius of the central black hole. These
observations confirm that TeV gamma-rays are emitted by extragalactic sources
other than blazars, where jets are not relativistically beamed towards the
observer.

{\bf The FSRQ 3C~279}. The Flat Spectrum Radio Quasar (FSRQ) 3C~279, with a
redshift of $z=0.536$, has recently been reported to be a TeV emitter by the
MAGIC Collaboration (Teshima et al. \cite{teshima07}). If confirmed, this is
the most distant AGN ever detected at TeV energies.

{\bf No starburst galaxy detection}. It is interesting to note here that the
two starburst galaxies NGC~253 and Arp~220 have been observed with HESS and
MAGIC, respectively. In the case of NGC~253, the new observations rule out a
previously claimed detection by CANGAROO-II (see Aharonian et al.
\cite{aharonian05_ngc253}). In the case of Arp~220, the nearest ultraluminous
infrared galaxy, the obtained upper limits are consistent with theoretical
expectations (Albert et al. \cite{albert07_arp220}). No starburst galaxy has
ever been firmly detected at VHE gamma rays.

{\bf Gamma-Ray Bursts (GRBs)}. Observations of long duration GRBs have been
conducted by the MAGIC telescope. In the case of GRB~050713a, since the
redshift of the GRB was not measured directly, the flux upper limit estimated
by MAGIC is still compatible with the assumption of an unbroken power-law
spectrum extending from a few hundred keV to energies above 175\,GeV (Albert et
al. \cite{albert06_grb050713a}). From 2005 April to 2006 February MAGIC
observed a total of 9 GRBs, obtaining upper limits in all cases. For the 4
bursts with measured redshift, the upper limits are compatible with a power law
extrapolation from the fluxes obtained at tens-to-hundred of keV (Albert et al.
\cite{albert07_grbs}). On the other hand, observations of short duration GRBs
have been possible with Milagro. Indeed, 17 GRBs lasting less than 5\,s have
fallen in the field of view of Milagro between 2000 January and 2006 December,
and upper limits have been obtained for all of them (Abdo et al.
\cite{abdo07_grbs}). Although the detection of long duration GRBs at high
redshift can be difficult due to absorption by the EBL, this is not the case
for the expected more nearby short duration GRBs. In any case, no detection has
been found for a relatively close burst at $z=0.225$ (see Abdo et al.
\cite{abdo07_grbs} for details).

\subsection{Galactic Sources}
\label{sect:galactic}
%\hspace{15pt}%                   %% preserved for Editor

{\bf Galactic Plane Surveys}. While serendipity in discovering extragalactic
sources would be rare, this is not the case for galactic ones, where galactic
plane surveys can play a major role in the discovery space. The most extreme
examples come from the galactic plane surveys conducted by HESS (Aharonian et
al. \cite{aharonian05_survey_science}; \cite{aharonian06_survey_apj}) and
Milagro (Abdo et al. \cite{abdo07_survey}).

The HESS survey of the inner Galaxy covers $\pm$30$^{\circ}$ in longitude and
$\pm$3$^{\circ}$ in latitude, with an average flux sensitivity of 2\% of the
Crab Nebula at energies above 200\,GeV. Up to 14 new sources were discovered,
12 being extended and 4 of them significantly elongated (Aharonian et al.
\cite{aharonian06_survey_apj}). Most of them are suggested to be supernova
remnants and/or pulsar wind nebulae, while two of them could be X-ray binaries
and about 6 remain unidentified (of which 3 have no suggested counterpart).

The Milagro galactic plane survey for declinations above $-7^{\circ}$ reveals
three multi-TeV sources apart from the already known emission from the Crab
Nebula. Their nature remains unknown. The most significant one, MGRO~J2019+37
(see also Abdo et al. \cite{abdo07_cygnus}), is clearly extended with a
diameter of the peak emission of $\sim1^{\circ}$ and contains several GeV
sources.

{\bf The Galactic Center}. Observations of the Galactic Center (GC) region with
HESS have revealed a point-like source in a position within 1${^\prime}$ of
Sgr~A$^*$ (Aharonian et al. \cite{aharonian04_sgra}). These observations are in
strong disagreement with previous ones reported by CANGAROO-II and show fluxes
somewhat below previous ones reported by Whipple. Although variability of the
supermassive black hole could be invoked, the TeV source has never been seen to
vary by any given instrument. MAGIC observations have confirmed the HESS
results: the flux, the spectrum and the lack of variability (Albert et al.
\cite{albert06_gc}). It should be noted that the TeV source at the Galactic
Center could also be the VHE counterpart of the SNR Sgr~A~East or of a recently
discovered pulsar wind nebula, in which cases no variability would be expected.
The possibility that the TeV emission has its origin in dark matter
annihilation processes has been recently studied, although the observed
power-law spectrum appears to be incompatible with the most conventional
scenarios (Aharonian et al. \cite{aharonian06_dark_matter}).

\begin{figure}
\centering
\resizebox{1.0\textwidth}{!}
{\includegraphics[]{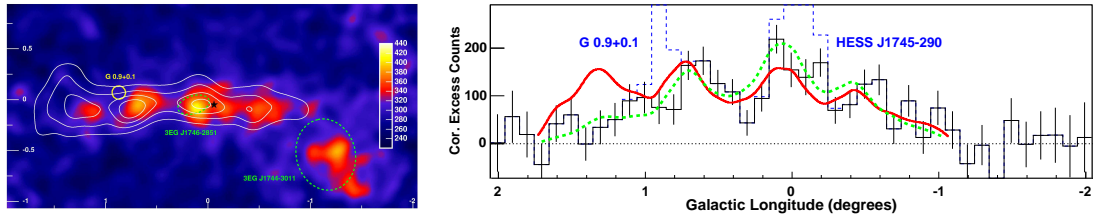}}
\caption{Left: VHE gamma-ray count map of the GC region after subtracting the GC source and SNR~G 0.9+0.1. A diffuse emission is clearly seen correlated with molecular gas traced by CS emission shown in white contours. Right: The red line shows the CS distribution, while the green dashed curve is the applied model to explain the data (see text). From Aharonian et al. (\cite{aharonian06_ridge}).}
\label{fig:ridge}
\end{figure}

{\bf The Galactic Center Ridge}. Deep HESS observations have revealed that
after subtracting the TeV contribution from the Galactic Center source and from
the SNR~G~0.9+0.1 (see below), a distribution of diffuse emission along the
plane is found (Fig.~\ref{fig:ridge}-left). This diffuse emission appears to be
well correlated with CS emission that traces molecular gas (a good target for
inelastic proton-proton interactions and neutral pion decay). Indeed, the
observed TeV morphology can be reproduced using a cosmic ray density
distribution and diffusion away from a central source of age $\sim10^4$ years
(Aharonian et al. \cite{aharonian06_ridge}).

%-----------------------------------------------------------------------------
%-----------------------------------------------------------------------------

{\bf SuperNova Remnants (SNR)}. SNR are thought to be the sites of cosmic ray
acceleration up to the knee, and TeV photons, which are not deflected by
galactic magnetic fields, can provide valuable information in this respect. Two
SNR displaying a shell at TeV energies have been found by HESS:
RX~J1713.7$-3946$ and RX~J0852$-$4622 (Aharonian et al.
\cite{aharonian04_rxj1713_nature} and \cite{aharonian05_rxj0852},
respectively). Interestingly the TeV shells follow nicely the synchrotron keV
emission seen by X-ray satellites. This suggests that a leptonic scenario could
be at work, where relativistic electrons produce the synchrotron X-ray emission
and IC up-scatter ambient photons to produce TeV emission. However, detailed
TeV spectra appear to be more compatible with a hadronic origin, where VHE
gamma-rays are produced by inelastic $pp$ interactions and $\pi^0$ decay
(Aharonian et al. \cite{aharonian06_rxj1713_spec},
\cite{aharonian07_rxj0852_spec}, \cite{aharonian07_rxj1713_100tev}).
Observations at lower energies with the {\it GLAST} satellite will allow to
unveil the nature of these VHE sources.

Two composite supernova remnants, SNR~G~0.9+0.1 and HESS~J1813$-$1718, have
also been found (Aharonian et al. \cite{aharonian05_snrg09},
\cite{aharonian06_survey_apj}). In the first case the VHE emission appears to
originate in the plerionic core and can be plausibly explained by IC scattering
of relativistic electrons. On the other hand, TeV emission has also been found
in two SNR near molecular clouds: HESS~J1834$-$087/W41 (Aharonian et al.
\cite{aharonian06_survey_apj}; Albert et al. \cite{albert06_hessj1834_w41}) and
MAGIC~J0616+225/IC~443 (Albert et al. \cite{albert07_ic443}). More precisely
the TeV emission appears to originate, specially in the second case, in the
interaction between the SNR and the molecular cloud, clearly suggesting a
hadronic scenario. Finally, it is worth to note here that a previously claimed
detection of SN~1006 has been ruled out by the upper limits obtained by HESS
(Aharonian et al. \cite{aharonian05_sn1006}).

{\bf Pulsar Wind Nebulae (PWN)}. The majority of the newly identified TeV
sources are PWN ($\sim$18 sources). In these objects a rapidly spinning neutron
star produces a relativistic wind of electrons that produce X-ray synchrotron
radiation. IC up-scattering of ambient photons, either CMB or from nearby
stars, produces the observed TeV emission. A good example of an extended TeV
PWN closely matching its X-ray emission is that of MSH~15$-$52 (Aharonian et
al. \cite{aharonian05_msh1552}). On the other hand, the PWN HESS~J1825$-$137
has been the first TeV source to display an energy-dependent morphology, in the
sense of a softening of the spectrum with increasing distance from the pulsar.
This favors a leptonic scenario where there is cooling of the electrons in the
nebula. Interestingly, the data is not compatible with a constant spin-down
power, and a higher injection power in the past is required (Aharonian et al.
\cite{aharonian06_hessj1825}).

{\bf Pulsars}. Although extensive searches have been conducted, no pulsed
emission from the Crab pulsar has been detected with MAGIC, constraining the
exponential cutoff energy to be below 27\,GeV, while for a super-exponential
the cutoff energy is below 60\,GeV (Albert et al. \cite{albert07_crab}). A
search for pulsed emission in 11 pulsars by HESS reveals that the VHE gamma-ray
production efficiency in young pulsars is less than 10$^{-4}$ of the pulsar
spin-down luminosity (Aharonian et al. \cite{aharonian07_pulsars}).

{\bf Galactic open clusters}. The extended source HESS~J1023$-$575 has been
found to be coincident with the young stellar cluster Westerlund~2, in the
well-known HII complex RCW49 (Aharonian et al. \cite{aharonian07_w2}).
Considered emission scenarios include emission from the colliding wind zone of
WR~20a, collective stellar winds, diffusive shock acceleration in the
wind-blown bubble itself, and supersonic winds breaking out into the ISM.

{\bf Unidentified sources}. Although some of the $\sim24$ still unidentified
sources have signatures of a PWN, there are good examples of sources with no
clear counterparts. The best example might be TeV~J2032+413, a steady and
extended ($FWHM\simeq14\pm3{^\prime}$) source displaying a hard spectrum
($\Gamma\simeq1.9$) discovered by HEGRA (Aharonian et al.
\cite{aharonian02_hegra}, \cite{aharonian05_hegra}). Although the apparent
absence of counterparts at lower energies suggested a dark accelerator, several
radio emitting X-ray sources have been recently discovered in its center of
gravity (see Paredes et al. \cite{paredes07} and references therein). More
recently, an extended ($FWHM\simeq12{^\prime}$) X-ray source matching the
position of TeV~J2032+413 has been revealed through deep {\it XMM-Newton}
observations (Horns et al. \cite{horns07}). In any case its nature, whether
hadronic or leptonic, remains unknown.

\section{X-ray/gamma-ray binaries}
\label{sect:binaries}
%\hspace{15pt}%                   %% preserved for Editor

Four X-ray binaries have been detected up to now at TeV energies:
PSR~B1259$-$63 (Aharonian et al. \cite{aharonian05_psrb1259}), LS~5039
(Aharonian et al. \cite{aharonian05_ls5039_science}), LS~I~+61~303 (Albert et
al. \cite{albert06_lsi_science}) and recently  Cygnus~X-1 (Albert et al.
\cite{albert07_cygx1}).

{\bf PSR~B1259$-$63} is composed of a 48~ms radio pulsar orbiting a B2\,Ve star
every 1237~d in a highly eccentric orbit with $e$=0.87. The pulsed radio
emission is not detected when the neutron star is behind the decretion disk of
the companion. The observed TeV spectrum can be fit with a simple power law,
but the lightcurve is puzzling when compared to previously available models.
Although it is clear that the interaction between the relativistic wind of the
young non-accreting pulsar and the polar wind and/or equatorial (but inclined)
decretion disk of the donor star plays a major role, the exact mechanism
producing the TeV emission is not known. Pure leptonic scenarios invoking IC
scattering have been put forward (Khangulyan et al. \cite{khangulyan07_psr}),
as well as  hadronic scenarios where TeV emission is produced by $\pi^0$ decay
and radio/X-ray emission by $\pi^{\pm}$ decay and IC scattering (Neronov \&
Chernyakova \cite{neronov07}). As it is the case for SNR, observations in the
GeV range by {\it GLAST} will provide the answer.

\begin{figure}
\centering
\resizebox{0.8\textwidth}{!}
{\includegraphics[]{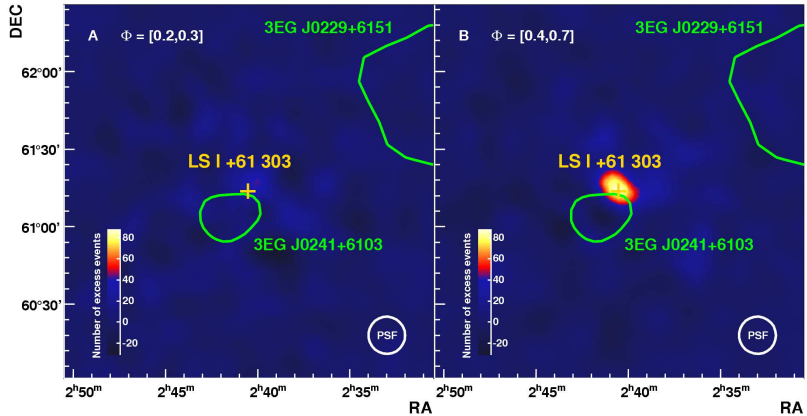}}
\caption{Smoothed VHE sky maps of LS~I~+61~303 in the orbital phase range
0.2--0.3 close to periastron passage (left) and in phase range 0.4--0.7 near
apastron (right). The variability is apparent. From Albert et al.
(\cite{albert06_lsi_science}).}
\label{fig:lsi_magic}
\end{figure}

{\bf LS~I~+61~303} is composed of a rapidly rotating early type B0\,Ve star
with a stable equatorial decretion disk and mass loss, and a compact object
with a mass between 1 and 4\,M$_\odot$ orbiting it every $\sim$26.5~d (see
Casares et al. \cite{casares05_lsi} and references therein). The TeV emission
detected by MAGIC is highly variable: upper limits have been found during
periastron passage and a peak occurs near apastron (Albert et al.
\cite{albert06_lsi_science}; see Fig.~\ref{fig:lsi_magic}). The spectrum can
always be fitted with a simple power law. It is worth to note that the VERITAS
Collaboration has recently confirmed the orbital TeV variability (Maier
\cite{maier07}). Massi et al. (\cite{massi04}) reported the discovery of an
extended jet-like and apparently precessing radio emitting structure at angular
extensions of 10--50~milliarcseconds. Due to the presence of (apparently
relativistic) radio emitting jets, LS~I~+61~303 was proposed to be a
microquasar. However, recent VLBA images obtained during a full orbital cycle
show a rotating elongated morphology that Dhawan et al. (\cite{dhawan06})
interpreted in the context of the interaction between the wind of the companion
and the relativistic wind of a young non-accreting ms pulsar, similarly as in
PSR~B1259$-$63 (see Dubus \cite{dubus06} for details on the model). The pulsed
radio emission would not be detected because of free-free absorption. However,
recent detailed simulations of this pulsar-wind interaction reveal a problem in
this scenario: to explain the observed GeV luminosity the spin-down power of
the putative pulsar should be so high that the wind of the companion could not
collimate the radio emitting particles (Romero et al. \cite{romero07}). The
nature of the source is thus still a matter of debate.

%-----------------------------------------------------------------------------
%-----------------------------------------------------------------------------

{\bf LS~5039} contains a compact object of unknown nature, with mass between
1.4 and 5\,M$_\odot$, orbiting every 3.9~days an ON6.5\,V((f)) donor in a
slightly eccentric orbit (Casares et al. \cite{casares05_ls}). The TeV flux of
this binary system is clearly periodic, with enhanced emission at inferior
conjunction of the compact object, suggesting that photon-photon absorption
(which has an angle dependent cross-section) within the system plays a major
role (Aharonian et al. \cite{aharonian06_ls5039_period}). However, the non-zero
flux observed at superior conjunction of the compact object and the lack of
variability at $\sim$200\,GeV argues against this simple interpretation, and
suggests that there may be an orbital phase-dependent electron acceleration
and/or that the TeV emission could be produced away from the compact object
(Khangulyan et al. \cite{khangulyan08_ls}). The detection of elongated
asymmetric emission in high-resolution radio images obtained with the VLBA and
the EVN was interpreted as evidence of its microquasar nature, and suggested
that the source was persistently producing mildly relativistic ejections with a
velocity of $\sim$0.15$c$ (Paredes et al. \cite{paredes00,paredes02}). Although
the X-ray spectra are compatible with those of accreting black holes while in
the so-called low/hard state (Bosch-Ramon et al. \cite{bosch05}), the radio
spectra are optically thin with a spectral index of $-$0.5 (Mart\'{\i} et al.
\cite{marti98}; Rib\'o et al. \cite{ribo99}). Theoretical modelling in the
microquasar scenario (Bosch-Ramon et al. \cite{bosch06}) has allowed to
reproduce the observed SED from radio to VHE gamma-rays (Paredes et al.
\cite{paredes06}). However, the lack of clear accretion signatures and the
similarities with the SEDs of PSR~1259$-$63 and LS~I~+61~303 has led other
authors to model its multi-wavelength emission using the scenario of wind
interactions (Dubus \cite{dubus06}). One of the predictions of this kind of
modelling is the periodic change in the direction and shape of the extended
radio morphology, as well as in the peak position of the radio core, depending
on the orbital phase. Therefore, high resolution VLBI observations can help to
unveil the real nature of the system.

{\bf Cygnus~X-1} is the most recent addition to the selected group of TeV
emitting X-ray binaries (Albert et al. \cite{albert07_cygx1}). This binary
system contains an O9.7\,Iab donor and an accreting black hole of at least
10\,M$_\odot$ orbiting it every 5.6 days in a circular orbit. It shows nearly
persistent radio emission, which sometimes has been resolved in jet-like
features that reveal its microquasar nature (Stirling et al.
\cite{stirling01}). A ring-like structure at arcminute scales has been
detected, showing the strong influence of the jet into its surrounding ISM
(Gallo et al. \cite{gallo05}). Cygnus~X-1 was detected by MAGIC only in a short
$\sim$80 minute time interval with a soft spectrum ($\Gamma\simeq$3.2)
extending up to $\sim$1\,TeV, while upper limits are obtained for the rest of
the $\sim$40~h observations. The position of the TeV emission is compatible
with that of Cygnus~X-1, excluding the ring-like structure. The TeV excess was
found at orbital phase 0.91, when the black hole is behind the star and
photon-photon absorption should be huge. For instance, Bednarek \& Giovanelli
(\cite{bednarek07}) computed the opacity to pair production for different
injection distances from the center of the massive star and angles of
propagation, finding that photons propagating through the intense stellar field
towards the observer would find in their way opacities of about 10 at 1\,TeV.
An origin in the jet of this microquasar and away from the compact object would
relax these values. The TeV detection took place during a particularly bright
hard X-ray state of Cygnus~X-1. Simultaneous hard X-ray observations by {\it
Swift}/BAT in the 15--50\,keV energy range reveal that the TeV excess was found
right before the onset of a hard X-ray peak. Observations one day later reveal
that no TeV excess was found during the maximum and decay phase of another hard
X-ray peak. Although one could speculate with the limited available data, more
simultaneous multi-wavelength observations are necessary to build reliable
models. In any case, this is the first experimental evidence of VHE emission
from a stellar-mass black hole, and therefore from a confirmed accreting X-ray
binary.

Finally, two X-ray binaries have been proposed as counterparts of unidentified
VHE sources: IGR~J16320$-$4751 for HESS~J1632$-$478 and IGR~J16358$-$4726 for
HESS~J1634$-$472 (Aharo\-nian et al. \cite{aharonian06_survey_apj}). However,
their extended TeV emission appears to rule out these associations.

\section{The future}
\label{sect:future}
%\hspace{15pt}%                   %% preserved for Editor

Regarding the near future, it is expected that online analyses will be able to
send alerts from Cherenkov telescopes to other astrophysical facilities. On the
other hand, given the success of instruments like HESS and MAGIC, in the longer
term the community wants to improve the sensitivity by up to a factor of
$\sim$10 and increase the energy range to cover from $\sim$30\,GeV to
$\sim$100\,TeV. In this context, a new Cherenkov telescope is being designed:
the CTA (Cherenkov Telescope Array)\footnote{\tt
http://www.mpi-hd.mpg.de/hfm/CTA/}. It is important to note here that a guest
observer program is planned for CTA.

\section{Conclusions}
\label{sect:conclusions}
%\hspace{15pt}%                   %% preserved for Editor

The third generation of IACTs has revealed several tens of sources in the TeV
sky. The nature of some of these sources was already known, and TeV
observations can help to unveil the leptonic or hadronic nature of the
accelerated particles, the ISM particle density, the magnetic fields, etc. On
the other hand, new types of sources have been discovered, boosting theory.
However, a lot of theoretical work is still needed to properly model some of
these sources. In addition, there are unidentified sources whose nature remains
unknown. New instruments like {\it GLAST} and CTA will allow to conduct
population studies and unveil the nature of some of the sources in the near
(and mid-term) future. In conclusion, the TeV sky is starting to shine and is
revealing new and very interesting laboratories for astrophysics.

\begin{acknowledgements}
The author acknowledges support by DGI of the Spanish  Ministerio de
Educaci\'on y Ciencia (MEC) under grant AYA2007-68034-C03-01 and FEDER funds.
He also acknowledges financial support from MEC through a \emph{Ram\'on y Cajal} fellowship.
This research has made use of the NASA's Astrophysics Data System Abstract 
Service, and of the SIMBAD database, operated at CDS, Strasbourg, France.
\end{acknowledgements}

\label{lastpage}

\end{document}